\documentclass[twocolumn,showpacs,showkeys,superscriptaddress]{revtex4}
\usepackage{graphicx}
\usepackage{amsmath}
\usepackage{amsfonts}
\usepackage{amssymb}
\usepackage{mathrsfs}

\def\openone{\leavevmode\hbox{\small1\kern-3.8pt\normalsize1}}
\def\N{\leavevmode\hbox{ Z \kern-8 pt\normalsize{Z}}}
\def\openone{\leavevmode\hbox{\small1\kern-3.8pt\normalsize1}}
\def\openJ{\leavevmode\hbox{J \kern-9.5pt\normalsize J}}
\def\openS{\leavevmode\hbox{ S \kern-9.3pt\normalsize S}}
\newcommand{\bb}{\begin{equation}}
\newcommand{\ee}{\end{equation}}
\newcommand{\eqb}{\begin{eqnarray}}
\newcommand{\eqf}{\end{eqnarray}}

\usepackage{color}

\begin{document}

\title{Gravitational electromotive force in magnetic reconnection around Schwarzschild black holes}
\author{Felipe A. Asenjo}
\email{felipe.asenjo@uai.cl}
\affiliation{Facultad de Ingenier\'{\i}a y Ciencias, Universidad Adolfo Ib\'a\~nez, Santiago 7941169, Chile.}
\author{Luca Comisso}
\email{luca.comisso@columbia.edu}
\affiliation{Department of Astronomy and Columbia Astrophysics Laboratory, Columbia University, New York, NY 10027, USA}


\begin{abstract}

We analytically explore the effects of the gravitational electromotive force on magnetic reconnection around Schwarzschild black holes through a generalized general-relativistic magnetohydrodynamic model that retains two-fluid effects. It is shown that the gravitational electromotive force can couple to collisionless two-fluid effects and drive magnetic reconnection. This is allowed by the departure from quasi-neutrality in curved spacetime, which is explicitly manifested as the emergence of an effective resistivity in Ohm's law. The  departure from quasi-neutrality is owed to different gravitational pulls experienced by separate parts of the current layer. This produces an enhancement of the reconnecion rate due to purely gravitational effects.

\end{abstract}


\maketitle


Magnetic fields are ubiquitous in the universe and they play a major role in a variety of astrophysical systems. At large scales, the behavior of highly conducting magnetized plasmas is well described by the equations of ideal magnetohydrodynamics (MHD), which impose significant constraints on the plasma dynamics. Indeed, an ideal MHD evolution implies the frozen-in condition and therefore the preservation of field line connectivity among fluid elements. This is a remarkably general result, which is valid in non-relativistic \cite{Newcomb}, special relativistic \cite{Newcomb,pegoraroEPJ,asenjoComissoCon}, as well as general relativistic \cite{asenjocomPRD17} plasmas.

On the other hand, at small spatial scales, physical effects beyond ideal MHD can break the frozen-in condition and allow for a topological rearrangement of the magnetic field configuration that occurs on time scales much faster than the global magnetic diffusion time. This process, known as magnetic reconnection \cite{YKJ_2010}, enables a rapid conversion of magnetic energy into plasma particle energy, and is generally believed to be the underlying mechanism that powers some of the most energetic astrophysical phenomena in the universe, such as solar and stellar flares \cite{Masuda94,Su13}, nonthermal signatures of pulsar wind nebulae \cite{Sironi11,Tavani11}, and gamma-ray flares in blazar jets \cite{Giannios09,Sironi15}.

Electrical resistivity due to Coulomb interactions between charged particles is the prototypical effect that can break the frozen-in condition and allow for the reconnection of magnetic field lines. This was indeed employed in many models of magnetic reconnection, from the pioneering Sweet-Parker model \cite{Sweet_1958,Parker_1957} to the more recent models of fast magnetic reconnection mediated by the plasmoid instability \cite{Huang_2010,ULS_2010,Comisso_2016,UzdLou_2016,Huang17,Comisso_2017}. Anomalous resistivity due to wave-particle interactions and scatterings off the turbulent fluctuations can also enable magnetic reconnection, and they have been considered as a possible agent of fast reconnection \cite{Ji04,Fox08,Che11}. Depending on the value of the classical/anomalous resistivity, other non-ideal effects can be even more important. For example, electron inertia effects are indeed known to permit nondissipative magnetic reconnection \cite{OP93,CafGrasso98,Comisso2013}, and in an analogous fashion, nongyrotropic electron pressure tensor effects can break the frozen-in constraint and sustain most of the reconnection electric field required for fast reconnection \cite{Hesse99,Shay07,Ari18}.

In relativistic plasmas, 
thermal effects proportional to the relativistic enthalpy density couple to the inertial effects, leading to an increase of the magnetic reconnection rate \cite{luca1,CApaper}. Furthermore, the Hall terms, that cannot cause magnetic reconnection per se in the nonrelativistic case, do allow for a change in the magnetic field line connectivity if there is a significant difference between the enthalpy density of the positively and negatively charged fluids constituting the plasma \cite{Kawazura17}. 
The situation is rendered even more complex in the presence of a strong gravitational field, as in the vicinity of compact objects like black holes. Several studies have predicted the formation of reconnection layers in the vicinity of black holes \cite{Koide2006,Karas2009,Penna2010,LyutMcK_2011,McKinney2012,Karas2012,Ball2018}, and the theoretical investigation of magnetic reconnection in curved spacetime has just started \cite{AsenjoComipaperKerr,CApaper}.

With this manuscript we intend to explore the effects of the gravitational electromotive force on magnetic reconnection in a curved spacetime around a black hole. In previous works \cite{AsenjoComipaperKerr,CApaper} the role of the radial gravitational force due to the black hole was not studied, as it requires a correct definition of the gravitational electromotive forces as well as understanding the influence of the charge density in curved spacetimes (see below). That the gravitational electromotive force contributes to magnetic reconnection was suggested by Koide \cite{Koide_2010}, without working out explicitly its quantitative effects on the reconnection rate.   
Here we focus on the simplest form of the gravitational field created by a black hole, i.e., a Schwarzschild black hole, and we calculate the reconnection rate due to the gravitational electromotive force.



In order to show that the gravitational field of a Schwarzschild black hole introduces new effects that are relevant for reconnection, we adopt a generalized version of the general-relativistic magnetohydrodynamic (GRMHD) equations \cite{Koide_2010,CApaper} which retain two-fluid effects that are neglected in the simpler single-fluid descriptions. In particular, we employ a set of equations \cite{CApaper} that describes electron-ion plasmas in the thermal-inertia regime \cite{kimura14,Lingam16}. This is the regime in which the thermal-inertial terms are larger than the Hall terms. Therefore, by taking into account the proper mass ratio between the positively and negatively charged particles, the same set of equations describes also pair plasmas, where the Hall terms vanish identically. 

The considered spacetime $x^\mu=(t,x^1,x^2,x^3)$ is characterized by a metric $g_{\mu\nu}$, where the line element is given by $d{s^2} = g_{\mu\nu} d{x^\mu} d{x^\nu}$.  Note that we choose units in which the speed of light $c$ is unity. 
The GRMHD equations deal with a single-fluid plasma model with proper enthalpy density $h = {n^2}({h_ + }/n_ + ^2 + {h_ - }/n_ - ^2)$, where  $n_\pm$ indicate the proper particle number density for the positively ($+$) and negatively ($-$) charged components fluids. Similarly, the enthalpy density $h_\pm$ of each charged fluid is specified with the corresponding subscript, and $n=n_+ + n_-$.
Furthermore, it is assumed that $\Delta h \ll h$, where $\Delta h = m{n^2}({h_ + }/{m_ + }n_ + ^2 - {h_-}/{m_-}n_-^2)/2$ is the difference between enthalpy densities of the fluids (with $m=m_+ + m_-$, and $m_\pm$ indicating the mass of the corresponding charged particle). It is also assumed the equation of state 
$h_\pm = m_\pm n_\pm {{{K_3}(m_\pm/{k_B}T_\pm)}}/{{{K_2}(m_\pm/{k_B}T_\pm)}}$ \cite{Chandra1938,Synge1957}, where $K_2$ and $K_3$ are the modified Bessel functions of the second kind of orders two and three,  $T_\pm$ are the temperatures of each fluid, and $k_B$ is the Boltzmann constant.

In this model the momentum equation that retains thermal-inertia effects is  \cite{Koide_2010,CApaper}
\begin{equation}\label{MomAp}
\nabla_\nu\left[h \left(U^\mu U^\nu+\frac{\xi}{4n^2e^2}J^\mu J^\nu\right)\right]=-\nabla^\mu p+J_\nu F^{\mu\nu}\, ,
\end{equation}
where $\nabla_\nu$ denotes the covariant derivative associated with the spacetime metric $g_{\mu\nu}$, $U^\mu$ is the plasma four-velocity, $J^\mu$ is the four-current density, and $F^{\mu\nu}$ is the electromagnetic field tensor. Furthermore, $p=p_+ + p_-$ indicates the proper plasma pressure, $e$ is the electron charge,  and $\xi=1-(\Delta\mu)^2$, with $\Delta\mu=({m_+-m_-})/({m_+ + m_-})$. Observe that $\xi\approx 4m_-/m_+$ for an electron-ion plasma, while $\xi=1$ for a pair plasma. 

Furthermore, the generalized Ohm's law in the thermal-inertial regime is  \cite{CApaper}
\begin{eqnarray}\label{InertialOhmAp}
U_\nu F^{\mu\nu} &=& \eta \left[ {J^\mu - \rho '_e   U^\mu} \right]  \nonumber\\
&+&\frac{\xi}{4e^2n}\nabla_\nu\left[\frac{h}{n}\left(U^\mu J^\nu+J^\mu U^\nu-\frac{\Delta\mu}{ne}J^\mu J^\nu\right)\right]  \, ,
\end{eqnarray}
where 
 $\rho '_e = -U_\nu J^\nu$ is the charge density observed by the local center-of-mass frame, and $\eta$ is the electrical resistivity, which is considered as a phenomenological parameter. Notice that, in comparison with the model equations of Ref.~\cite{CApaper}, we are considering a plasma where the thermal energy excange rate between the two fluids is negligible \cite{Koide_2010}, i.e., the redistribution coefficient of the thermalized energy to the positively and negatively charged fluids is zero.

The plasma dynamics is completed by the continuity equation
\begin{equation}\label{contAp}
\nabla_\nu\left(n U^\nu\right)=0\, ,
\end{equation}
and Maxwell's equations
\begin{equation}\label{MaxGeneralCurvedAp}
\nabla_\nu F^{\mu\nu}=J^\mu \, , \qquad \nabla_\nu F^{*\mu\nu}=0 \, ,
\end{equation}
where $F^{*\mu\nu}$ is the dual of the electromagnetic field tensor.

To explicitly display the gravitational effects in the above plasma model in a familiar fashion, we write the previous equations in the $3+1$ formalism \cite{TM_82,Thorne86,Zhang89}. In such form, that  spacetime curvature effects become aparent in a set of vectorial equations. 
For a Schwarzschild background, with spherical geometry, the line element becomes 
\begin{equation}\label{lineeleme}
d{s^2} =  - {\alpha ^2}d{t^2} + h_1^2 dr^2 + h_2^2 d\theta^2+h_3^2 d\phi^2\, ,
\end{equation}
with 
$\alpha=\sqrt{1-{2 r_s}/{r}}$, $h_1={1}/{\alpha}$, $h_2=r$, and $h_3=r\sin\theta$.
Here, $\alpha$ is known as the lapse function, $r$ is the radial distance to the black hole, $r_s$ is the half of the Schwarzschild radius (hereafter $G=1=c$), $0\leq\theta\leq\pi$, and $0\leq\phi\leq\ 2\pi$. In order to properly describe the plasma dynamics, it is also useful to re-write the plasma vectorial equations by introducing a locally nonrotating frame called ``zero-angular-momentum-observer'' (ZAMO) frame \cite{Bardeen_1972,Koide_2010,CApaper,AsenjoComipaperKerr}, which introduces a locally Minkowskian spacetime in where the line element \eqref{lineeleme} can be written as $d{s^2} =  - d{{\hat t}^2} + \sum\limits_{i = 1}^3 {{{(d{{\hat x}^i})}^2}}$,
where $d\hat t = \alpha dt$ and $d{{\hat x}^i} = {h_i}d{x^i}$. In the following, quantities observed in the ZAMO frame are denoted with hats.

We first consider the continuity equation \eqref{contAp}, which can be rewritten in the ZAMO frame as \cite{Koide_2010, CApaper} 
\begin{equation}\label{contZAMO}
\frac{\partial(\gamma n)}{\partial t}+\frac{\alpha}{r^2\sin\theta}\sum_j\frac{\partial}{\partial x^j}\left(\frac{r^2\sin\theta}{h_j}\gamma n\hat v^j\right)=0\, ,
\end{equation}
where $\hat v$ is the velocity  in the ZAMO frame, and $\gamma  = {(1 - {\hat v^2})^{ - 1/2}}$ is the  Lorentz factor (we use latin indices  for space components).
We also consider the spatial components of the generalized momentum equation \eqref{MomAp}, which lead to the dynamical equation
\begin{eqnarray}\label{momZAMO}
\frac{\partial \hat P^i}{\partial t}&=&-\frac{\alpha}{r^2\sin\theta}\sum_j\frac{\partial}{\partial x^j}\left(\frac{r^2\sin\theta}{h_j}\hat T^{ij}\right)\nonumber\\
&&-(\epsilon+\gamma \rho)\frac{1}{h_i}\frac{\partial\alpha}{\partial x^i}
+\sum_j \alpha\left[G_{ij}\hat T^{ij}-G_{ji}\hat T^{jj}\right]\, ,
\end{eqnarray}
where
\begin{equation}\label{PT1}
\hat P^i = h\gamma^2 \hat v^i+\frac{h \xi}{4 n^2e^2}\hat J^i\hat J^0+\sum_{j,k}\varepsilon_{ijk}\hat E_j \hat B_k \, ,
\end{equation}
\begin{equation}\label{}
\epsilon = h\gamma^2+\frac{h \xi}{4e^2n^2}(\hat J^0)^2-p-\rho\gamma+\frac{1}{2}\left(\hat B^2+\hat E^2\right) \, ,
\end{equation}
and 
\begin{eqnarray}\label{PT2}
\hat T^{ij}&=&p\delta^{ij}+h\gamma^2\hat v^i\hat v^j+\frac{h \xi}{4e^2n^2}\hat J^i \hat J^j\nonumber\\
&&+\frac{1}{2}\left(\hat B^2+\hat E^2\right)\delta^{ij}-\hat B_i\hat B_j-\hat E_i\hat E_j \, .
\end{eqnarray} 
Here, $\hat J^0$ is the separation of charge density  while $\hat J^i$ is the current density, both observed in the ZAMO frame. It is the main goal of this work to  show (below) that $\hat J^0$ affects the magnetic reconnection process by the gravitational electromotive force. 
Besides, it is important to notice that $\hat J^0$ is related to the invariant  $\rho_e'=-U_\mu J^\mu$.
We also specify that $\hat E_j$ and $\hat B_j$ are the electric and magnetic fields measured in the ZAMO frame, $G_{ij}=-({1}/{h_ih_j})({\partial h_i}/{\partial x^j})$, and $\varepsilon_{ijk}$ is the Levi-Civita symbol.

For the spatial components of the generalized Ohm's law \eqref{InertialOhmAp}, in the ZAMO frame we have
\begin{eqnarray}\label{OhmZAMO}
&&\frac{\xi}{en}\frac{\partial}{\partial t}\left[\frac{h\gamma}{4en} \left(\hat J^i+\hat J^0\hat v^i\right)\right]=-\frac{h \xi \gamma\hat J^0}{2e^2n^2 h_i}\frac{\partial\alpha}{\partial x^i}\nonumber\\
&&\quad\qquad-\frac{\alpha}{en r^2\sin\theta}\sum_j\frac{\partial}{\partial x^j}\left(\frac{r^2\sin\theta}{h_j}\hat K^{ij}\right)\nonumber\\
&&\qquad\quad+\frac{\alpha}{en}\sum_j \left(G_{ij}\hat K^{ij}-G_{ji}\hat K^{jj}\right)\nonumber\\
&&\qquad+\alpha\gamma \hat F_{i0}+\alpha\gamma\hat v^j\hat F_{ij}-\alpha\eta \left(\hat J^i - \rho '_e  \gamma \hat v^i \right)\, ,
\end{eqnarray}
where $\hat K^{ij}=({h \xi \gamma}/{4en})(\hat v^i\hat J^j+\hat v^j\hat J^i)$. Similarly, the temporal component of Eq. \eqref{InertialOhmAp} becomes \cite{Koide_2010}
\begin{eqnarray}\label{OhmZAMOtime}
&&\frac{\xi}{2en}\frac{\partial}{\partial t}\left(\frac{h\gamma \hat J^0}{en}\right)=-\frac{h \xi \gamma}{4e^2n^2}\sum_j\frac{1}{h_j}\frac{\partial\alpha}{\partial x^j}\left(\hat J^j+\hat J^0\hat v^j\right)\nonumber\\
&&\quad-\frac{\alpha}{en r^2\sin\theta}\sum_j\frac{\partial}{\partial x^j}\left(\frac{r^2\sin\theta}{h_j}\frac{h \xi \gamma}{4en}\left[\hat J^j+\hat J^0\hat v^j\right]\right)\nonumber\\
&&\qquad+\alpha\gamma \hat v^j\hat F_{j0}-\alpha\eta \left(\hat J^0 - \rho '_e  \gamma \right)\, .
\end{eqnarray}

Finally, we rewrite Maxwell's equations \eqref{MaxGeneralCurvedAp} in the ZAMO frame.  These are
\begin{equation}\label{Maxw1ZAMO}
\sum_j\frac{\partial}{\partial x^j}\left(\frac{r^2\sin\theta}{\alpha h_j}\hat B_j\right)=0\, ,
\end{equation}
\begin{equation}\label{Maxw3ZAMO}
\frac{\alpha}{r^2\sin\theta}\sum_j\frac{\partial}{\partial x^j}\left(\frac{r^2\sin\theta}{\alpha h_j}\hat E_j\right)=\hat J^0\, ,
\end{equation}
\begin{eqnarray}\label{Maxw2ZAMO}
\alpha\hat J^i+\frac{\partial \hat E_i}{\partial t}=\frac{\alpha h_i}{r^2\sin\theta}\sum_{j,k}\varepsilon^{ijk}\frac{\partial}{\partial x^j} \left(\alpha h_k  \hat B_k\right) \, ,
\end{eqnarray}
\begin{eqnarray}\label{Maxw4ZAMO}
\frac{\partial \hat B_i}{\partial t}=\frac{-\alpha h_i}{r^2\sin\theta}\sum_{j,k}\varepsilon^{ijk}\frac{\partial}{\partial x^j} \left( \alpha h_k  \hat E_k\right) \, .
\end{eqnarray}


The gravitational field of a Schwarzschild black hole introduces effects in the generalized Ohm's law \eqref{OhmZAMO} that can be seen as effective electric fields. In particular, terms with the form $G_{ij}\hat K^{ij}$ and $G_{ji}\hat K^{jj}$ in  Eq.~\eqref{OhmZAMO} can introduce effective resistivities of the order $(h \xi/4en)(\partial_j h_i/h_i h_j)$, in where both the gravitational field and the thermal-inertial effects are important. However, as we will see below, in the simplest possible geometry for the reconnection layer, these both terms vanish. 
On the other hand, as noticed by Koide in Ref.~\cite{Koide_2010}, the term proportional to $\hat J^0 ({\partial_i\alpha}/h_i)$ in Eq. \eqref{OhmZAMO} produces a radial contribution to the generalized Ohm's law that can be interpreted as an effective electric field, as long as $\hat J^0$ does not vanish. Therefore, in this work we analyze this possibility, showing that a reconnection layer around a Schwarzschild black hole allows a solution in which the separation of charge $\hat J^0$ is finite, and that in this case the electromotive force due to gravity can drive magnetic reconnection.

\begin{figure}[]
\begin{center}
\includegraphics[width=8.2cm]{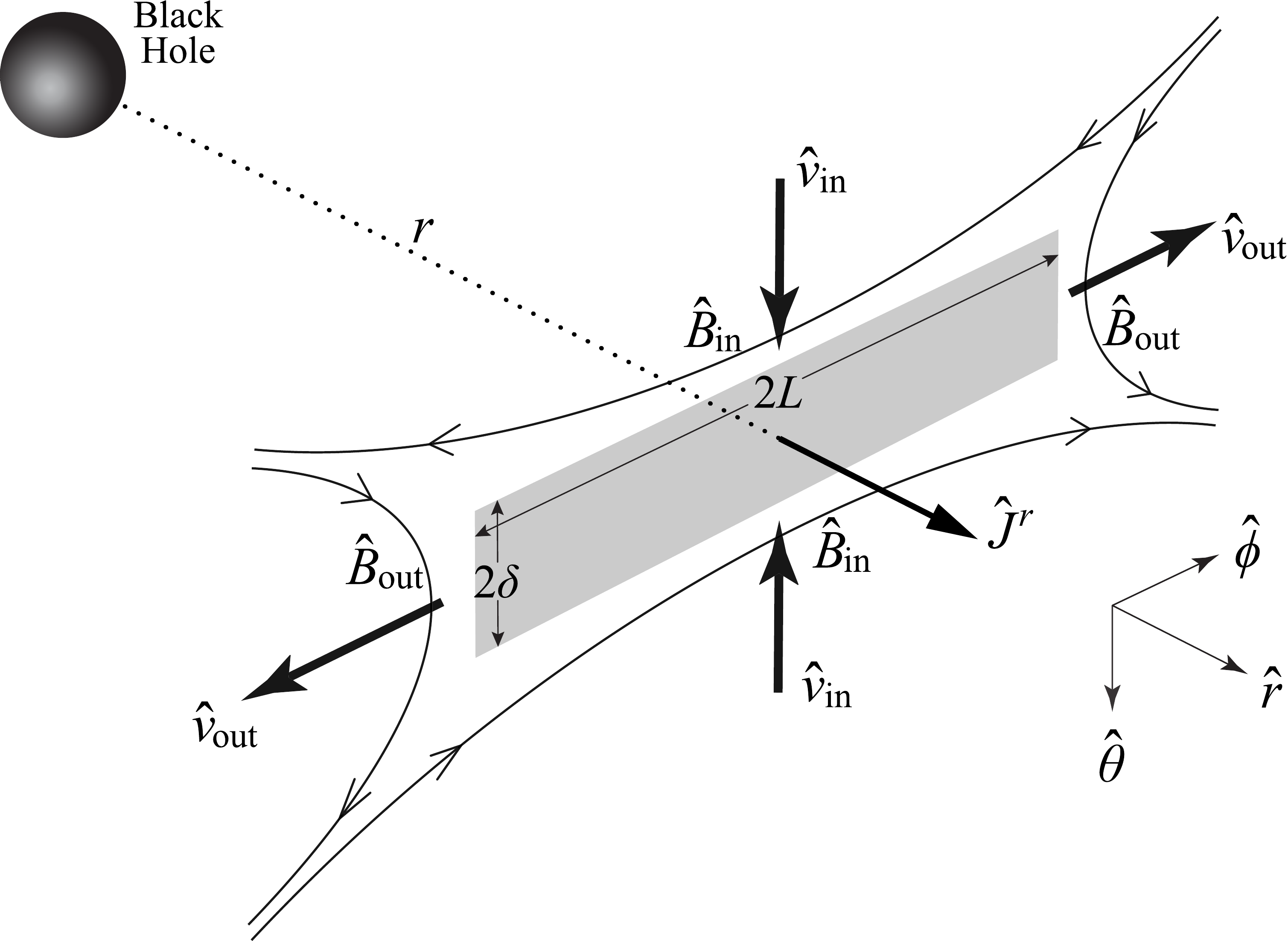}
\end{center}
\caption{Sketch of a magnetic reconnection layer showing the studied configuration. The shaded gray area represents the magnetic diffusion region.}
\label{fig1}
\end{figure}

Without loss of generality, let us assume that the reconnection layer is at $\theta=\pi/2$ at some given distance $r$. We consider a quasi--two--dimensional reconnection layer having characteristic length $L$ and width $\delta$ such that $\delta\ll L$. The length $L$ is in the $\phi$-direction, while the width $\delta$ is in the $\theta$-direction, as depicted in Fig.~\ref{fig1}. We also assume that the layer is not close to the black hole, $\delta\ll L \ll r$. This model allow us to study magnetic reconnection using a Sweet-Parker-like approach for a plasma that is supported against the black hole gravity \cite{abramo,Koide2011,Tursunov2016}, as for the model we investigated in Kerr curved spacetime \cite{AsenjoComipaperKerr,CApaper}. We also assume that the radial plasma velocity is null or negligible, i.e. $\hat v^r=0$, and that in the diffusion region $\hat J^\theta=0$ and $\hat J^\phi=0$. Then, it is important to observe that $\rho_e'=-U_\mu J^\mu=\gamma \hat J^0\neq 0$, in general \cite{Koide_2010}.
Furthermore, the reconnecting magnetic field has magnitude $\hat B_{\rm{in}}$ in 
$\phi$--direction (with no radial component in the reconnection layer), while the electric field is in the radial direction (see Fig.~\ref{fig1}).

Under the above assumptions, we can readily calculate the outflow velocity of the plasma accelerated through the reconnection channel. This plasma outflow is in $\phi$-direction along the neutral line.
By using the momentum equation \eqref{momZAMO}  we find
\begin{eqnarray}\label{momZAMO2}
\sum_j\frac{\partial}{\partial x^j}\left(\frac{r^2\sin\theta}{h_j}\hat T^{\phi j}\right) = 0 \, ,
\end{eqnarray}
as other terms indentically vanish along the $\phi$--direction. The  solution for this equation  is $T^{\phi\phi}=0$. 
Taking the tensor \eqref{PT2} along the neutral line, and using that $p \approx \hat B_{\rm{in}}^2/2  \approx h/4$ in the relativistic regime \cite{AsenjoComipaperKerr,CApaper},  we can readily find that the outflow plasma velocity satisfies $\gamma_{\rm{out}}\hat v_{\rm{out}}\approx {1}/{\sqrt{2}}$.

Similarly, we can estimate other relevant quantities for this reconnection layer configuration. From the divergenceless equation \eqref{Maxw1ZAMO}, the outflow magnetic field in the $\theta$-direction is
\begin{equation}\label{bo}
\left.\hat B_\theta \right|_{\rm{out}}\approx \frac{\delta}{L} \hat B_{\rm{in}}\, .
\end{equation}
On the other hand, using the continuity equation \eqref{contZAMO} for the Schwarzschild geometry, the inflow plasma velocity can be written as
\begin{equation}\label{vin}
 {\gamma_{\rm{in}}\hat v_{\rm{in}}}\approx\frac{\delta}{L}\,  {\gamma_{\rm{out}}\hat v_{\rm{out}}}\, .
\end{equation}
Besides, from Eq.~\eqref{Maxw2ZAMO} we obtain that the radial current density at the $X$ point is simply 
\begin{equation}\label{jr}
\left. \hat J^r\right |_{\rm{X}}\approx \frac{\hat B_{\rm{in}}}{\delta}\, .
\end{equation}
The results \eqref{bo}, \eqref{vin} and \eqref{jr} are equivalent to those pertaining relativistic plasmas in flat spacetimes \cite{lyu,luca1}. The explanation for this is  the chosen configuration around the Schwarzschild black hole. The simple geometry studied here, with the invoked assumptions,  implies that no gravitational effects appear in the momentum equation or Maxwell's equations when they are evaluated in the reconnecion layer. As we shall see now, all the gravitational effects appear in the generalized Ohm's law.


We focus on the spatial part of the generalized Ohm's law \eqref{OhmZAMO} along $r$-direction. For our geometry, in the current sheet this equation becomes
\begin{eqnarray}\label{OhmZAMO999}
&& \frac{\alpha}{e n r^2\sin\theta}\sum_j \frac{\partial}{\partial x^j}\left(\frac{r^2\sin\theta}{h_j} \frac{h \xi \gamma}{4 e n} \hat v^j \hat J^r\right) + \frac{h \xi  \alpha \gamma \hat J_0}{2e^2 n^2}\frac{\partial\alpha}{\partial r} = \nonumber\\
&&+ \alpha \gamma \hat E_r-\alpha\gamma \hat v^\theta \hat B_\phi+\alpha\gamma \hat v^\phi \hat B_\theta -\alpha \eta \hat J^r\, .
\end{eqnarray}
We evaluate this equation in the inflow point, where the inflow plasma velocity is in the $\theta$-direction and the term proportional to the resistivity is negligible. Thus, we get
\begin{equation}\label{electricinpoint}
\left.\hat E_r\right|_{\rm{in}}\approx \hat v_{\rm{in}} \hat B_{\rm{in}}+\left.\frac{h \xi \hat J_0}{2e^2 n^2}\frac{r_s}{\alpha r^2}\right|_{\rm{in}}\, ,
\end{equation}
where we have used that ${\partial_r\alpha}={r_s}/{\alpha r^2}$. Here, we have  neglected the non-linear terms and considered $\gamma_{\rm{in}} \approx 1$, in agreement with the results of Refs.~\cite{AsenjoComipaperKerr,CApaper}. 
We can also evaluate Eq.~\eqref{OhmZAMO999}  at the $X$-point (where the plasma velocity vanishes), obtaining
\begin{equation}\label{electricinpointx}
\left.\hat E_r\right|_{X}\approx (\eta+\Lambda) \hat J^r+\left.\frac{h \xi \hat J_0}{2e^2 n^2}\frac{r_s}{\alpha r^2}\right|_{X}\, ,
\end{equation}
where we have introduced the effective  relativitistic collisionless resistivity \cite{luca1}
\begin{equation}
\Lambda=\frac{h \xi }{4 e^2 n^2 L}\, .
\end{equation}
Both results \eqref{electricinpoint} and \eqref{electricinpointx} reduce to those of Ref.~\cite{luca1} in the flat spacetime limit  $r_s\rightarrow 0$.

As there is no quasi--neutrality, with $\hat J^0$ different from zero, the electric fields  $\hat E_r |_{\rm{in}}$
and $\hat E_r |_{\rm{X}}$ are not equal. This is due to the presence of different gravitational gradients at the inflow and $X$ points. The radial distance of the inflow point $r|_{\rm{in}}$ is related to the radial distance $r$ of the $X$-point  by $r|_{\rm{in}}\approx r+{\delta^2}/({8r})$, where $r|_X\equiv r$, and thereby the two points experience slightly different gravitational pulls.
We can obtain the difference between the electric field at the inflow and $X$ points by using Eq.~\eqref{Maxw3ZAMO}. By integration among these two points in the current layer, and the radial distance of the inflow point, we get
\begin{equation}\label{diverelectriJ000}
\left.\hat E_r\right|_{\rm{in}}-\left.\hat E_r\right|_{X}\approx\frac{\delta^2}{8\alpha r} \hat J^0\, ,
\end{equation}
where the lapse function must be evaluated at the distance $r$ of the $X$-point.

What remains to be done is to obtain a relation between the current density and $\hat J^0$. This can be achieved through the temporal part  of the generalized Ohm's law, namely Eq.~\eqref{OhmZAMOtime}.
We can use that  $\rho_e'=\gamma \hat J^0$ is an invariant to calculate $\hat J^0$ by evaluation of Eq.~\eqref{OhmZAMOtimeb}   at the outflow point.
Thereby, assuming that the variations of the current density are neglegible in this geometry compared to the gravitational gradient, i.e.,  neglecting the divergence of the current density with respect to the gradient of the lapse function projected along the current
\begin{equation}
\frac{\alpha}{r^2}\frac{\partial}{\partial r}\left(\alpha r^2 \hat J^r\right)\ll \alpha\frac{\partial\alpha}{\partial r}\hat J^r\, ,
\end{equation}
then from Eq.~\eqref{OhmZAMOtime} evaluated in the outflow point we obtain
\begin{eqnarray}\label{OhmZAMOtimeb}
0&\approx& -\left.\frac{\gamma\Lambda L r_s}{r^2} \hat J^r\right|_{\rm{out}} -\left.  \frac{\alpha\gamma\Lambda L}{r}\frac{\partial}{\partial\phi} \left(\hat v^\phi \hat J^0\right) \right|_{\rm{out}}\nonumber\\
&&+\left.\alpha \eta  \gamma^2\hat v^2 \hat J^0\right|_{\rm{out}}\, ,
\end{eqnarray}
where we have used that $1-\gamma^2=-\gamma^2 \hat v^2$. As $\hat J^0$ decreases to the $X$-point, 
the previous equation can be solved for $\hat J^0$ to finally get
\begin{equation}\label{equationforJ0}
\hat J^0\approx \frac{2\Lambda L \chi r_s}{\alpha r^2\left(\eta+\Lambda\right)}\hat J^r\, ,
\end{equation}
where $\chi=1-{L^2}/({4r^2})-{r_s L^2}/({8\alpha^2 r^3})$, and  we have used the radial distance of the outflow point $r|_{\rm{out}}\approx r+{L^2}/({8r})$ in terms of the radial distance $r$ of the $X$-point. Notice that the separation of charge $\hat J^0$ is only relevant  in curved spacetimes, as it vanishes  when $r_s\rightarrow 0$.

Finally, using the above equations,
we can obtain the reconnetion rate for this configuration. In order to preserve the validity of our result, we restrict ourselves to a plasma sufficiently far from the black hole, $r_s\ll r$. In this case, the reconnection rate becomes simply
\begin{equation}\label{reconnfinal}
\hat v_{\rm{in}}\approx \left(\frac{1}{S}+\frac{\Lambda}{L}\right)^{1/2}\left[1+\frac{\Lambda L^2   r_s}{8\alpha^2  r^3 (\eta+\Lambda)}\right]\, ,
\end{equation}
where $S=L/\eta\gg 1$ is the relativistic Lundquist number.

The result  \eqref{reconnfinal} shows that the gravitational electromotive  force increases the reconnection rate due to purely the gravitational attraction of the  Schwarzschild black hole, compared to the MHD limit  $\hat v_{\rm{in}}\approx {S}^{-1/2}$  (when $\Lambda=0$). In the flat spacetime limit, $r_s\rightarrow 0$, we recover the reconnetion rates for a special relativistic pair plasmas $\hat v_{\rm{in}}\approx ({1}/{S}+{\Lambda}/{L})^{1/2}$ studied in Ref.~\cite{luca1}. 




The physical mechanism for the increase of the reconnection rate due to gravity is straightforward to understand. The gravitational force (due to gradients of $\alpha$) at the inflow point is along the radial direction at an angle $\theta\approx \pi/2-\delta/(2r)$. This is the force proportional to $\hat J^0 (r_s/\alpha r^2)|_{\rm{in}}$ that appears in Eq.~\eqref{electricinpoint}. On the other hand, the gravitational force  that the plasma experiences at the $X$--point is also along the radial direction but now at an angle $\theta=\pi/2$. Anew, this force is proportional to the term $\hat J^0 (r_s/\alpha r^2)|_{\rm{X}}$ in Eq.~\eqref{electricinpointx}. 
These two gradient forces point in radial direction  at different angles, implying the existence of a net force antiparallel to the $\theta$--direction, along  the plane of the reconnection layer. Therefore, the net force pushes the plasma toward the $X$--point, producing an increase of the reconnection rate.  

In case in which  the difference of gravitational forces between the inflow and $X$ points is neglected, then the  plasma can be considered as quasi--neutral, with $\hat J^0=0$. This is the case of the analyses presented Refs.~\cite{CApaper,AsenjoComipaperKerr}, where quasi--neutral plasma were studied around Kerr black holes, and only the curvature due to spacetime rotation was considered.
However, if the most general case for the simplest gravitational effect produced by any compact object is considered into the study of magnetic reconnection in the surrounding plasma, a deviation from quasi--neutrality is expected.

Finally, the reconnection rate \eqref{reconnfinal} explicity display the importance of taking into consideration the collisionless effects. Those effects are the ones coupled to gravity. In particular, the difference  of the reconnection rate \eqref{reconnfinal} in the limit $S \to \infty$ and its flat spacetime counterpart $\hat v_{\rm{in}} \approx \sqrt{\Lambda/L}$, is proportional to
\begin{equation}
\left.\frac{\hat v_{\rm{in}}}{\sqrt{\Lambda/L}}-1\right|_{S \to \infty}\propto\left(\frac{h}{m_-  n}\right)^{1/2}\left(\frac{d_e}{16}\right)\left(\frac{L}{r}\right)^2\left(\frac{2r_s}{r}\right)\, ,
\end{equation}
for pair plasmas, and proportional to 
\begin{equation}
\left.\frac{\hat v_{\rm{in}}}{\sqrt{\Lambda/L}}-1\right|_{S \to \infty}\propto\left(\frac{h}{m_+ n}\right)^{1/2}\left(\frac{d_e}{8}\right)\left(\frac{L}{r}\right)^2\left(\frac{2r_s}{r}\right)\, ,
\end{equation}
for ion--electron plasmas (here $d_e=\lambda_e/L$ is the dimensionless electron inertial length, with $\lambda_e$ indicating the electron skin depth). Both results show that reconnection rates are larger in plasmas around Schwarzschild black holes, depending on the size of the black hole $\propto 2r_s/r$, and on the geometry of the current sheet $\propto L/r$. Nevertheless, the reconnection rate for pair plasmas is larger according to the fact that positrons contribute as the electrons to the effective relativitistic collisionless resistivity $\Lambda$. 

The presented results complete the theoretical analysis of magnetic reconnection in curved spacetime initiated in Refs.~\cite{AsenjoComipaperKerr,CApaper}. In this way, we have shown that spacetime curvature effects (gravitational pull or rotation) form an intrinsic part of magnetic reconnection processes in astrophysical plasmas around compact objects.
Future high-resolution numerical simulations with general relativistic codes should be able to extend the predictions of the analytic theory to more complex scenarios, as asymmetric reconnection layers, strong field inhomogeneities in all three spatial directions, and non-steady reconnection processes.

\begin{acknowledgments}
 F.A.A. thanks Fondecyt-Chile  Grant No. 1180139.
\end{acknowledgments}

\end{document}